\documentclass[conference]{IEEEtran}

\usepackage{microtype}
\usepackage{graphicx}
\usepackage{booktabs}
\usepackage{caption}
\usepackage{subcaption}

\usepackage[hidelinks]{hyperref}

\usepackage[inline]{enumitem}
\usepackage{amsmath}
\usepackage{amssymb}
\usepackage{mathtools}
\usepackage{amsthm}
\usepackage{bbm}
\usepackage{multirow}
\usepackage{multicol}

\usepackage[capitalize,noabbrev]{cleveref}
\usepackage{booktabs}
\usepackage{afterpage}

\usepackage{changepage}
\usepackage{xcolor}
\usepackage{algorithm}
\usepackage{algorithmic}

\DeclareMathOperator*{\argmax}{arg\,max}
\DeclareMathOperator*{\argmin}{arg\,min}
\newcommand{\VertTab}[1]{\rotatebox{75}{#1}}

\theoremstyle{plain}

\theoremstyle{definition}

\theoremstyle{remark}

\begin{document}

\title{Bad Citrus: Reducing Adversarial Costs with Model Distances}

\author{%Anonymous Authors
\IEEEauthorblockN{Giorgio Severi}
\IEEEauthorblockA{\textit{Northeastern University}}
\and
\IEEEauthorblockN{Will Pearce}
\IEEEauthorblockA{\textit{Nvidia}}
\and
\IEEEauthorblockN{Alina Oprea}
\IEEEauthorblockA{\textit{Northeastern University}}
}
\date{}

\maketitle
\thispagestyle{plain}
\pagestyle{plain}

\begin{abstract}
Recent work by Jia et al.~\cite{jia_zest_2022}, showed the possibility of effectively computing pairwise model distances in weight space, using a model explanation technique known as LIME.
This method requires query-only access to the two models under examination.
We argue this insight can be leveraged by an adversary to reduce the net cost (number of queries) of launching an evasion campaign against a deployed model.
We show that there is a strong negative correlation between the success rate of adversarial transfer and the distance between the victim model and the surrogate used to generate the evasive samples.
Thus, we propose and evaluate a method to reduce adversarial costs by finding the closest surrogate model for adversarial transfer.
\end{abstract}

% \begin{IEEEkeywords}
% \end{IEEEkeywords}

\section{Introduction}

Evasion attacks~\cite{szegedy_intriguing_2014}, often referred to as adversarial examples, have been a strong focus of machine learning (ML) researchers for quite some time now.
Despite the large body of work on the subject, launching evasion campaigns, that is finding multiple adversarial examples for a given deployed model, remains a non-trivial task, especially when the adversary is given only query access to the victim.
There are two main strategies to tackle this issue:
\begin{enumerate*}[label=(\roman*)]
  \item using an attack method based on a zeroth-order optimization technique such as AutoZOOM~\cite{tu_autozoom_2019} and HopSkipJump~\cite{chen_hopskipjumpattack_2020};
  \item crafting the adversarial examples on a local surrogate (proxy) model, using a gradient-based approach, and transferring~\cite{demontis_why_2019} the generated evasive points to the victim.
\end{enumerate*}

As with most practical applications of adversarial machine learning, both strategies come with specific trade-offs.
Gradient-free methods, while effective, are limited by their strict threat model.
They generally require either a large number of queries to the victim model for each adversarial example, (and) or they tend to create more distorted points than their gradient-based counterparts.
Adversarial transfer, on the other hand, allows the attacker to leverage the full power of gradient-based methods, and enable them to generate a large number of  evasive samples without inducing high query volumes.
However, the rate of success of transferred samples is not guaranteed to be satisfactory, leading to potentially failing attack campaigns.
%and is generally higher if the attacker uses a proxy model similar to the actual victim.
Therefore, an adversary willing to launch an evasion campaign against a deployed ML classifier has to carefully consider costs related to query API usage, and ensure they are not identified by anomaly detection systems deployed to monitor incoming queries to the victim model.

Given this cost landscape for the adversary, any technique aimed at increasing the rate of successful transfer of adversarial examples crafted locally on proxy models would result in a direct decrease in the costs of running an attack campaign.
Based on this insight, we argue that the recently proposed "Zest of LIME" paper by Jia et al.~\cite{jia_zest_2022}, offers an intriguing new approach for the adversary to minimize their costs.
\cite{jia_zest_2022} proposes a methodology to compute distances between pairs of models given only access to the their outputs using Local Interpretable Model-agnostic Explanations (LIME)~\cite{ribeiro_why_2016}, which builds local linear models based on the answer of the target model to specific queries.
A small set of representative points ($N=128$ in the paper) is used to build $N$ linear regression models approximating the classifier around the chosen points, forming its LIME representation (signature).
These representations are then used to compute the distance between two target classifiers.

While this process does require the adversary to pay a cost for the queries used to estimate the LIME representation, that cost is only paid once -- the learned representation can be saved and re-used at any time.
Moreover, the adversary can progressively collect LIME representations for an abundance of potential proxy models, either by downloading them from model hubs or by manually training diverse models locally, and, over time, build a library of representations to compare against new victims.
This would induce an economy of scale effect due to which, the bigger the adversarial library becomes, the easier it becomes for the adversary to find good proxy models for generating evasive samples.

\section{Background and Related Works}

In this section we provide a short introduction to the concepts behind Zest distances and the main approaches currently used for carrying out black-box evasion attacks.

\subsection{Model Distances with Zest}

Local Interpretable Model-agnostic Explanations (LIME)~\cite{ribeiro_why_2016} is a model interpretability approach focused on training surrogate linear models to locally approximate the predictions of the model under analysis.
LIME is architecture agnostic, and only requires query access to the model making it viable for use with a remote target behind an API.

Recent work by Jia et al.~\cite{jia_zest_2022} proposes a new approach to estimate the similarity between different models based on LIME, called Zest.
Zest offers a variety of advantages with respect to previous model comparison methods.
First, it is generally easier to apply than direct weights comparison, as the latter requires both full access to the weight matrices, and that the models under scrutiny share the same architecture.
It is also less susceptible to inconsistencies related to the selection of representative inputs than other methods based on comparing model predictions. 
At its core, Zest samples $N$ images form the training set to use as reference inputs and then generates $N$ corresponding LIME linear regression models by perturbing super-pixels representing continuous pixel patches, and querying the target model with the perturbed inputs.
These local models are then aggregated into \emph{signatures} over which a distance metric such as Cosine similarity, or the $L_{1/2/\infty}$ norm of the difference, is applied.

\subsection{Black-box Attacks}

Many common adversarial example generation techniques, \cite{goodfellow_explaining_2015, carlini_towards_2017, madry_towards_2018}, are developed in a white-box scenario, where the adversary can compute gradients on the loss of the model with respect to its weights. 
They generally lead to effective attacks, able to alter the prediction of the model with extremely limited perturbation budgets, which are very useful in estimating the robustness of both existing models and proposed defensive systems.
However, they are generally not applicable in realistic scenarios where the adversary's only interface with the victim model is through an API designed by the model owner, which usually returns only the classifier's output and is provided under a pay-per-query model.
Efforts to circumvent these limitations led to the development of two main strategies: gradient-free methods, and transfer attacks.

Examples of the first class include methods based on generating adversarial examples through zeroth-order optimization techniques, such as ZOO~\cite{chen_zoo_2017} and AutoZOOM~\cite{tu_autozoom_2019}.
Another commonly used technique, HopSkipJump~\cite{chen_hopskipjumpattack_2020}, focuses on estimating the direction of the gradient by analyzing the outputs of the model in the proximity of the decision boundary.
Square Attack~\cite{andriushchenko_square_2020}, on the other hand, approaches the problem by adapting a randomized search scheme where each perturbation is confined to a small square of pixels.
These methods are generally characterized by a rather large amount of queries to the victim model for each adversarial example generated.

The second, studied by~\cite{demontis_why_2019, liu_delving_2017, mao_transfer_2022}, revolves around using one or multiple local proxy (or surrogate) models to compute a set of adversarial examples, and then using these locally generated points to attack the victim model.
The use of proxy models implies that the adversary is not limited in which technique to use to generate the evasive samples, and can leverage the full power of gradient based methods.
This approach allows the attacker to minimize the number of queries to the victim model, as they are free to generate a vast quantity of evasive points locally without repeatedly generating large query volumes.
However, the rate at which the generated samples transfer successfully to the victim is highly variable with the proxy model used in the generation phase, and generally hard to predict.
\section{Threat Model}

In this work we are interested in a realistic scenario where the adversary wishes to craft evasive samples for a victim model while having limited, query-only, access to it, often referred to in literature as \emph{black-box}.
This means that the adversary can send requests to the victim model, and retrieve the classifier output scores for each query.
The number of queries the adversary can make is strictly limited by their resources, as they will have to pay a fixed cost to run each query.
This is an extremely common situation, as most deployed machine learning systems provide access through a payed API system, which generally returns only the output scores for each query\footnote{In the rarer cases in which the target model returns only categorical labels, LIME would not be applicable. The adversary would have to fall-back to gradient-free methods.}.
While there are a multitude of works in the adversarial ML literature exploring a variety of other threat models, we argue that this represents one of the most realistic and wide-spread scenarios.

\subsection{Problem statement}

Let us assume the adversary is in control of a number $n$ of different classification models, proxies $\{ p_1, ..., p_n \}$, trained on a similar data distribution as the victim model $f$.
We consider a transfer attack successful if the perturbed data point, generated using only information gathered through the local proxy model, induces a mis-classification in the target model, $f$.
Therefore, the adversary's objective $\mathcal{A}$ is to generate a set of adversarial examples $A = \{ a_1, ..., a_m \}$, starting from clean test points $D_t = \{ (x_1, y_1), ..., (x_m, y_m) \} $,  such that the largest possible number of them are evasive for $f$: 
\begin{equation}
\begin{gathered}
 \mathcal{A}_j = \sum_{i=1}^m \mathbbm{1} ( f(a_i) \neq y_i ) \\
 \mathcal{A} = \argmax_{p \in P} \mathcal{A}_p
\end{gathered}
\end{equation}

In this work we formulate and empirically analyze the following hypotheses:
\begin{itemize}
    \setlength\itemsep{1em}
    % \item [\textbf{H1}] Adversarial examples computed on models in $P$  will transfer to $m$ with different success rates. In particular, we expect models of a similar architecture
    % to produce adversarial examples that transfer with higher success rates. 
    \item [(H1)] Pairs of models $(p_j, f)$ with similar architectures will show, on average, lower Zest distances.
    \item [(H2)] There is a negative correlation between the Zest distance of a pair $(p_j, f)$ and the successful transfer rate of adversarial examples from $p_j$ to $f$.
\end{itemize}
Testing these hypotheses is critical in determining if Zest distances between models can be used for reducing the cost of black-box adversarial attacks. (H1) is informative in determining the relationships between models trained on similar architectures.
(H2) implies that an adversary can directly leverage Zest distances as a source of information to select the best possible surrogate when targeting a black-box model.

\section{Methodology}

The process followed by the attacker to increase the cost-effectiveness of their campaigns is very simple, and summarized in \Cref{alg:bad_lemon}.
It starts with acquiring a large number of models trained for a similar task as the victim model.
For each collected model, they would compute the respective LIME representation and store it for later reuse.
Note that this process does not have to be temporally bounded.
The adversary can continue to accumulate useful models, progressively updating their collection of LIME representations.
After having acquired a sizeable portfolio of potential proxy models, the adversary can generate the LIME representation of the victim model, which will only require access to the inputs (selected by the adversary) and the victim logits, and so can be obtained for many deployed victim models.

Once in possession of the necessary LIME representations, the adversary can proceed to compute pairwise distances between their portfolio of proxy models and the target, using the method proposed in~\cite{jia_zest_2022}.
Finally, the adversary can select the proxy model with the minimal distance from the target, and craft a large variety of adversarial examples using strong gradient-based methods such as Projected Gradient Descent (PGD)~\cite{madry_towards_2018}.

\begin{algorithm}
\footnotesize

\caption{Bad Citrus}\label{alg:bad_lemon}
\begin{algorithmic}

\REQUIRE $D_N$ a subset of training points of size $N$
\REQUIRE $D_t$ a set of valid points for which the adversary wishes to generate evasive variants
\REQUIRE $\{ p_1, ..., p_n \}$ proxy models and a target model $f$
\ENSURE $A \gets$ adversarial examples from best proxy model

\vspace{5pt}

\STATE \textcolor{darkgray}{// Compute LIME signatures for each model}
\FOR{$j = 1; \ j < n; \ j++;$}
\STATE $l_j \gets \mbox{LIME}(p_j, D_N)$
\ENDFOR

\vspace{5pt}

\STATE $l_f \gets \mbox{LIME}(f, D_N)$
\STATE $dist \gets \{ \}$
\vspace{5pt}

\STATE \textcolor{darkgray}{// Compute the Zest distance between each pair $(p_j, f)$}
\FOR{$j = 1; \ j < n; \ j++;$}
\STATE $dist_j \gets \mbox{Zest}(l_j, l_f)$
\ENDFOR
\STATE $p \gets \argmin (dist)$

\vspace{5pt}

\STATE $A \gets \mbox{generateAdvEx}(p, D_t)$

\end{algorithmic}
\end{algorithm}

\section{Experimental Evaluation}

In this section we empirically evaluate our hypotheses and highlight the negative correlation between successful adversarial transfer and Zest distance.

\subsection{Experimental setup}

For our evaluations, we use the well known CIFAR-10 dataset\footnote{\url{https://www.cs.toronto.edu/~kriz/cifar.html}}, containing 60,000 images of 32x32 pixels.
We use a set of 13 pre-trained CIFAR-10 models, leveraging the PyTorch implementations provided by~\cite{phan_huyvnphanpytorch_cifar10_2021}, including the following types of architecture: DenseNet, ResNet, MobileNet V2, GoogleNet, Inception V3 and VGG (with batch normalization).

\begin{figure}
\vskip 0.2in
\centerline{\includegraphics[width=1\linewidth]{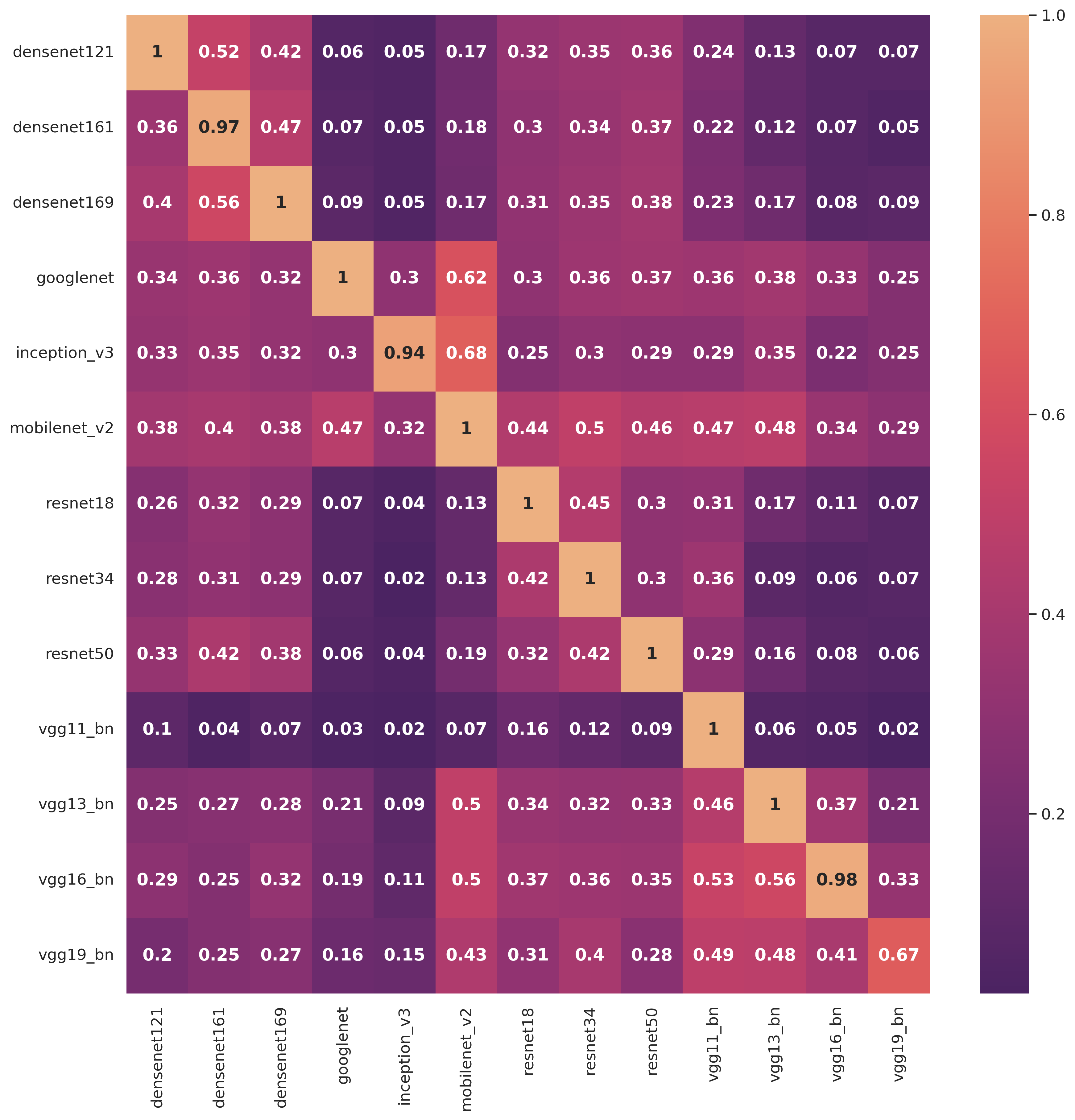}}
\caption{Rate of successful adversarial transfer between all pairs of CIFAR-10 models.}
\label{transfer_all}
\vskip -0.2in
\end{figure}

For all models, we use PGD\footnote{We use the open source implementation from the Adversarial Robustness Toolbox~\url{https://github.com/Trusted-AI/adversarial-robustness-toolbox}} to craft untargeted adversarial examples.
Unless otherwise noted, we select an $l_\infty$ budget $\epsilon = 0.1$, a step size of 0.02, and 5 random restarts.
This setup results in $> 90\%$ local attack success in all cases with the exception of VGG-19\footnote{We are primarily interested in showing the correlation between transfer rate and model distance. An attacker can invest resources tuning the attack parameters to craft even more effective samples.}.
The successful adversarial transfer rate for all pairs of CIFAR-10 models is shown in \Cref{transfer_all}.
% In all experiments, we also compute the Zest distance using both the $L_\infty$ and Cosine metrics.
The same subset of representative points was used for all models when computing their LIME signature, and all the adversarial examples were generated on the same subset of 100 images from the CIFAR-10 test set, originally classified correctly by all models.
% All experiments are run on a single Nvidia Titan X GPU with 12GB of VRAM.

\subsection{Model Distances}
\label{sec:distance}

We computed the Zest distance among all possible pairs of our 13 CIFAR-10 models, using the default parameters from \cite{jia_zest_2022}, under both the $L_\infty$ and Cosine metrics. 
The default value of the number of points used to compute the LIME signatures in Zest is $N=128$, and each point is perturbed 1000 times.
As highlighted in \Cref{tab:families_closest}, the Cosine metric appears more effective in associating models coming from the same architecture family, finding a closest model of the same type for 9/10 models.
The $L_\infty$ metric, on the other hand, tends to consider models from different architectures as closest, especially in the case of ResNets, where all three models are associated with models from the DenseNet family.

The full set of distances is reported in 
\Cref{tab:all_both_distances128}, \Cref{tab:all_both_distances64}, \Cref{tab:all_both_distances32} for $N=128$, $N=64$, and $N=32$ respectively\footnote{All distances reported are not normalized - this is not an issue as we are interested in their relative ordering and deltas, rather then their absolute value.}.
We measure the distances for lower values of $N$, as smaller $N$ values reduce the costs, in terms of number of queries, for the adversary to compute the LIME signature of the target model.
The relative order of measured distances for $N=128$ and $N=64$ appear quite aligned, while the measurements for $N=32$ tend to differ more.
This is in line with what was reported by~\cite{jia_zest_2022}, who mentioned that the standard deviation of measurements increases with smaller reference sets.

\begin{table}
\centering
\footnotesize
\caption{Closest model for each target in an architecture family by Zest $L_\infty$ and Cosine distances, $N=128$.}
\vskip 0.15in
\label{tab:families_closest}
\begin{tabular}{llll}
\textbf{} & \textbf{Target} & \textbf{Closest} & \textbf{Distance} \\
\toprule
$\mathbf{L_\infty}$& densenet121 & densenet161 & 2.5259 \\
          & densenet161 & densenet169 & 2.4783 \\
          & densenet169 & resnet50 & 2.4038 \\
          & resnet18 & densenet169 & 2.6694 \\
          & resnet34 & densenet121 & 2.5648 \\
          & resnet50 & densenet169 & 2.4038 \\
          & vgg11\_bn & inception\_v3 & 2.9164 \\
          & vgg13\_bn & vgg16\_bn & 2.6872 \\
          & vgg16\_bn & vgg13\_bn & 2.6872 \\
          & vgg19\_bn & vgg13\_bn & 4.0845 \\
\midrule
\textbf{Cosine}    & densenet121 & densenet161 & 0.1405 \\
          & densenet161 & densenet121 & 0.1405 \\
          & densenet169 & densenet121 & 0.1406 \\
          & resnet18 & resnet34 & 0.1545 \\
          & resnet34 & resnet18 & 0.1545 \\
          & resnet50 & densenet169 & 0.1441 \\
          & vgg11\_bn & vgg13\_bn & 0.1602 \\
          & vgg13\_bn & vgg16\_bn & 0.1295 \\
          & vgg16\_bn & vgg13\_bn & 0.1295 \\
          & vgg19\_bn & vgg16\_bn & 0.1615 \\

\end{tabular}
\vskip -0.1in
\end{table}

\subsection{Zest-Transfer Correlation}

\begin{figure*}
\footnotesize
    \centering
    \subfloat[DenseNet161, Cosine\label{dens161_cos_corr}]{%
      \includegraphics[width=0.4\linewidth]{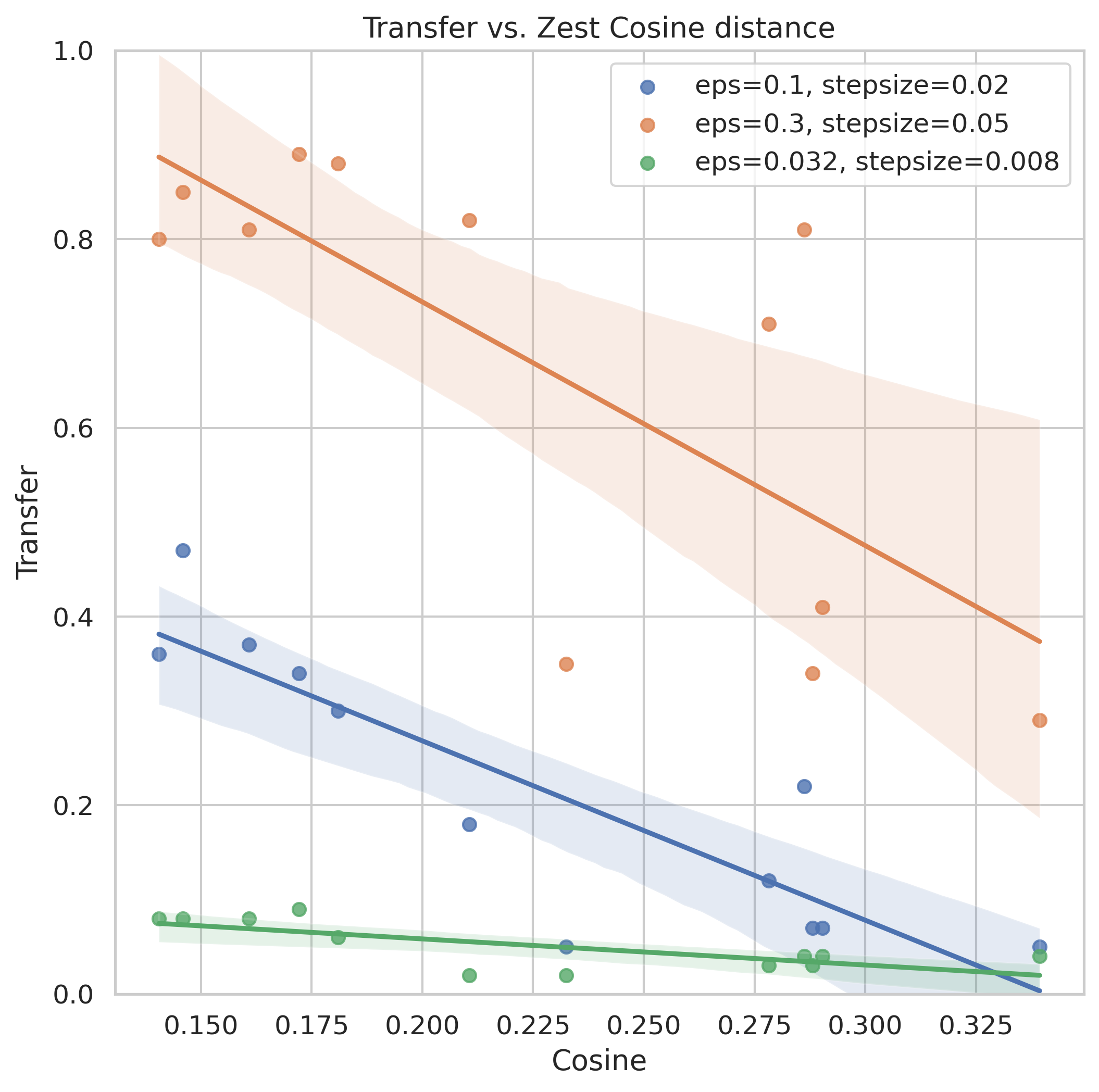}}
    \subfloat[DenseNet161, $L_\infty$\label{dens161_linf_corr}]{%
      \includegraphics[width=0.4\linewidth]{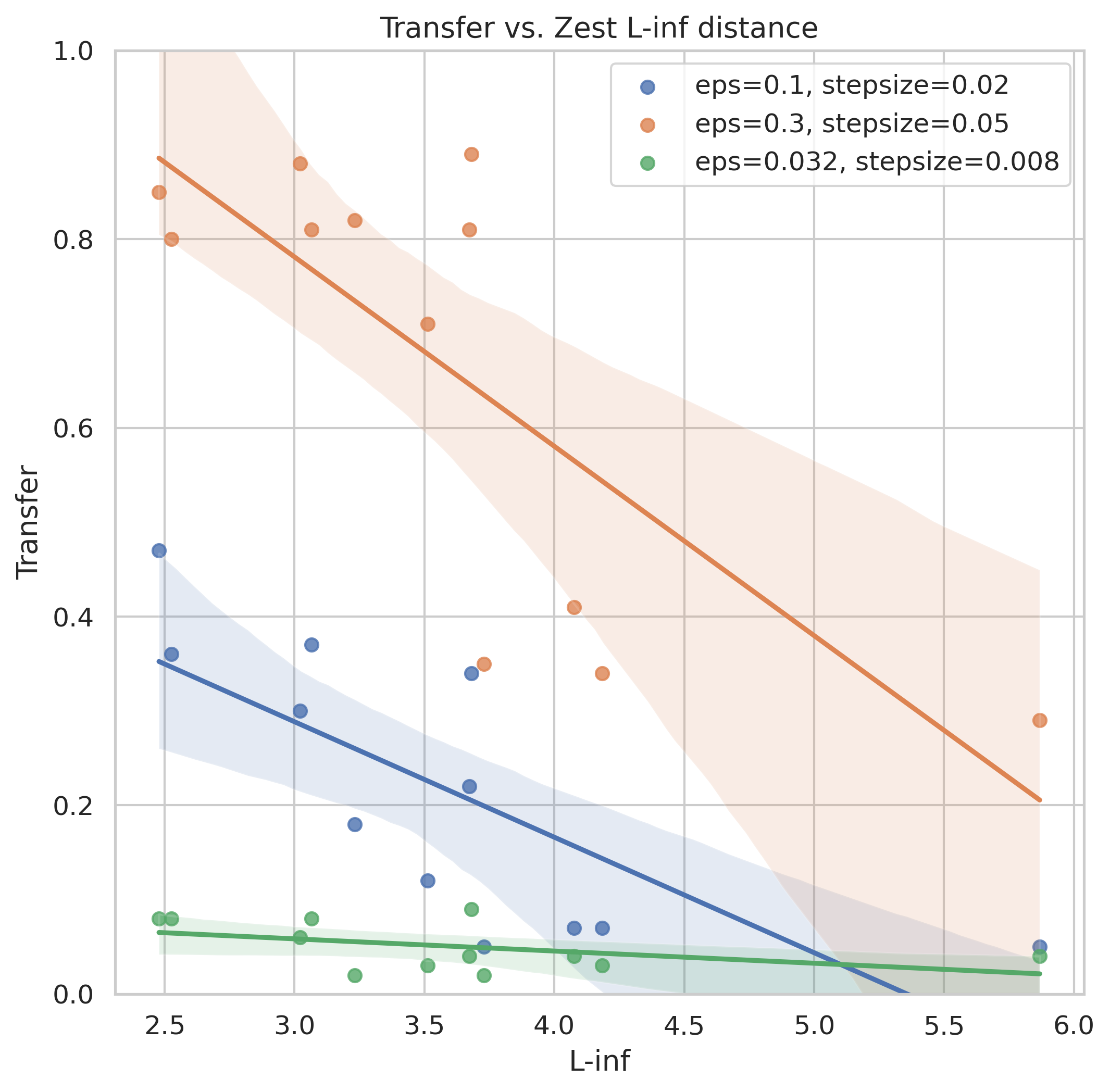}}
    \\
    \vspace{10pt}
    \subfloat[MLaaS model, Cosine\label{mlaas_128v32_cosine}]{%
      \includegraphics[width=0.4\linewidth]{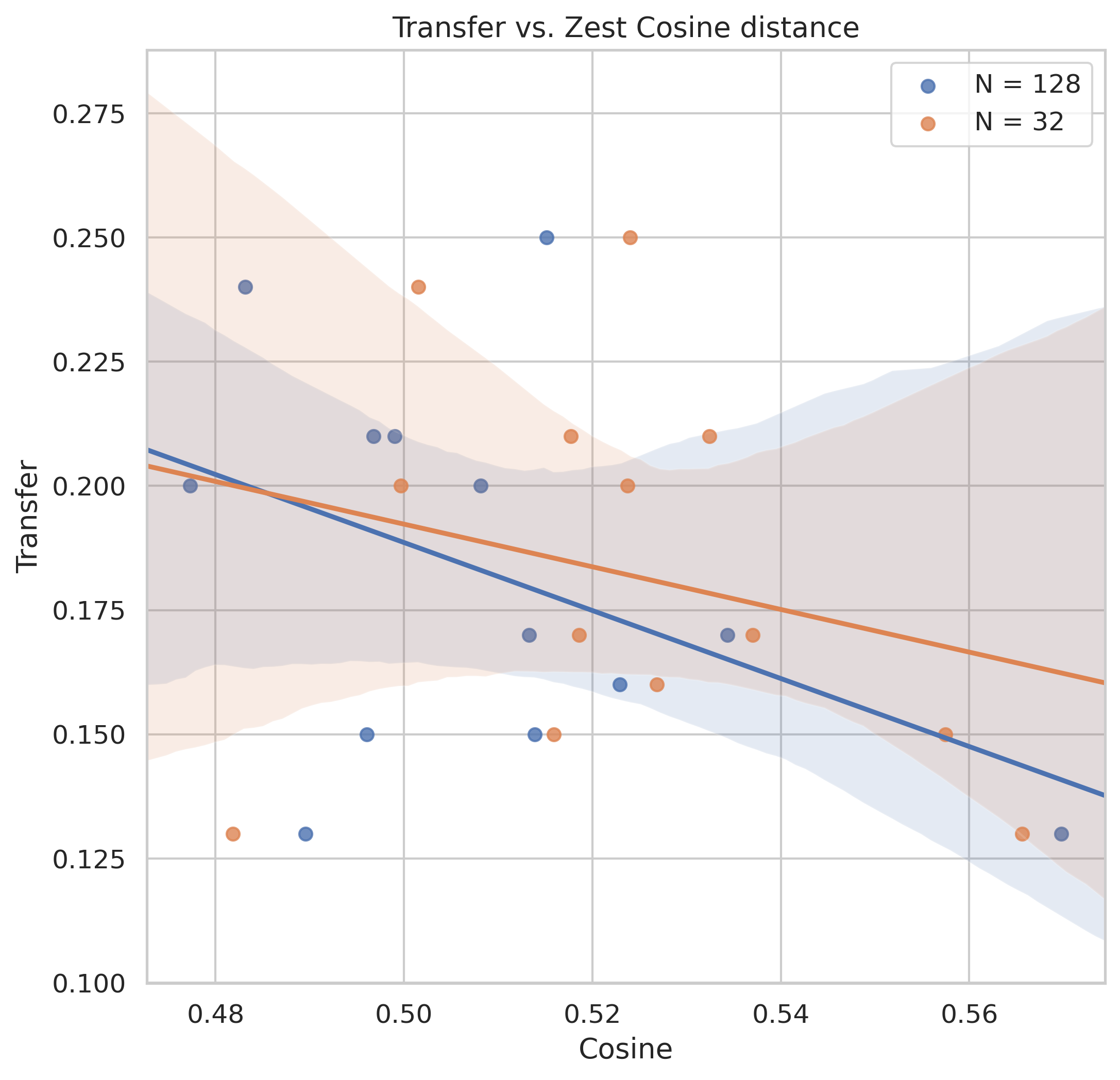}}
    \subfloat[MLaaS model, $L_\infty$\label{mlaas_128v32_linf}]{%
      \includegraphics[width=0.4\linewidth]{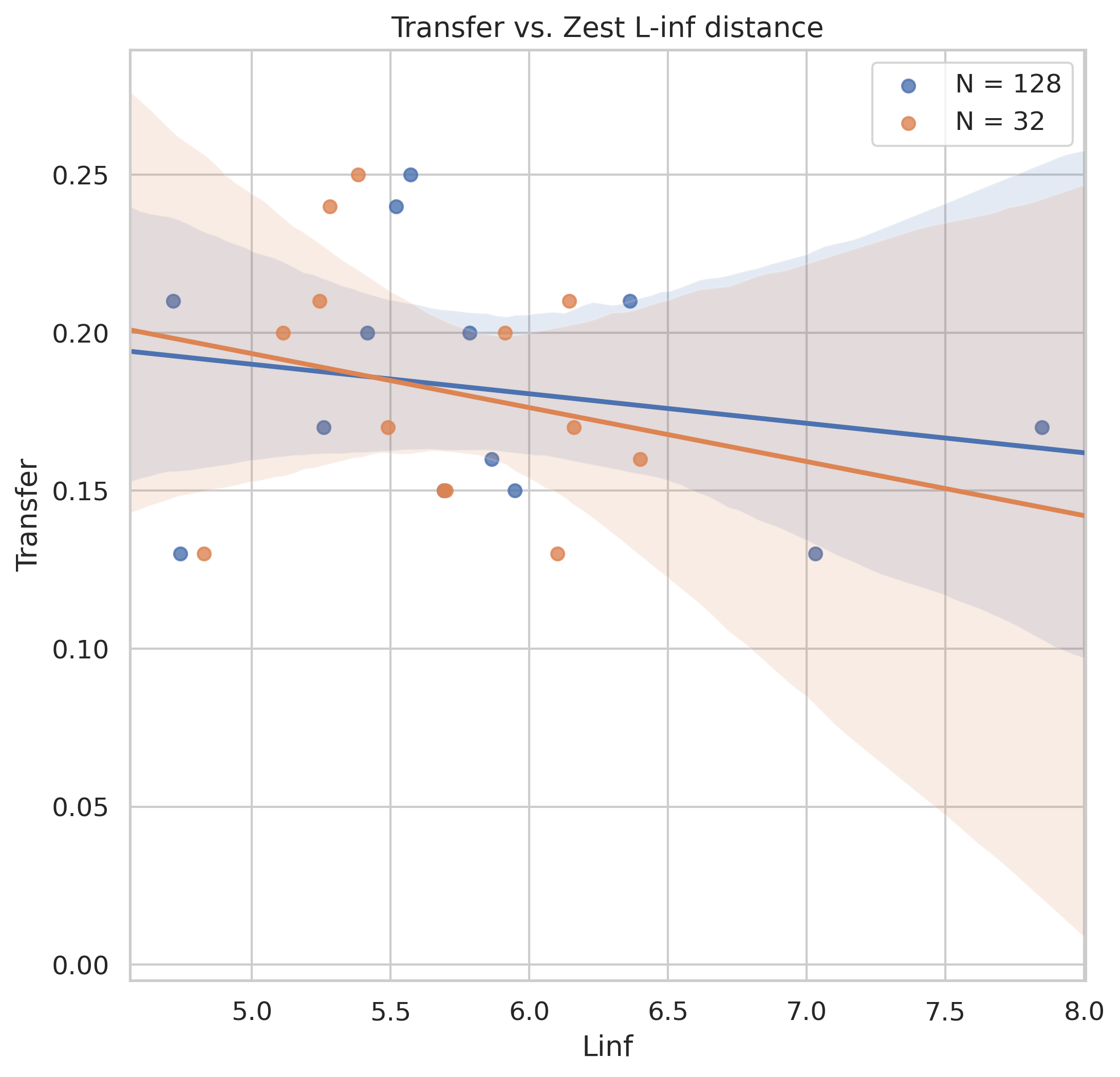}}
    \\
    \caption{Rate of successful adversarial transfer plotted against Zest distance. The top row shows results for a DenseNet161 model, while the bottom row shows results for the MLaaS classifier.}
\vskip -0.15in
\end{figure*}

To evaluate (H2), we measured the Zest distance between a randomly selected victim model (DenseNet 161), and all the remaining pre-trained models. 
We then run the PGD attack with three different $l_\infty$ perturbation budgets, $\epsilon \in \{0.3, 0.1, 0.032\}$, corresponding to a relatively large, medium and small visible perturbation.
\Cref{dens161_cos_corr} and \Cref{dens161_linf_corr} show the rate of successful adversarial transfer, plotted against the value of the Zest distance under the $L_\infty$ and Cosine metrics, for $N=128$.
We find a strong negative correlation between the Zest distance and the rate of successful transfer, for both distance metrics, and all attack budget.
In particular, for $\epsilon = 0.1$, we estimate a Pearson correlation coefficient of -0.746 and -0.871 for $L_\infty$ and Cosine respectively.

Selecting the middle value of $\epsilon = 0.1$ and repeating the experiment for all the 13 CIFAR-10 models, we obtain an average correlation coefficient of -0.594 for Cosine and -0.439 for $L_\infty$. 
We observe that the average correlation is significantly impacted by the the very low adversarial transfer success towards VGG-11 from all other proxy models, highlighted by \Cref{transfer_all}.
The higher average correlation, together with the observations from \Cref{sec:distance}, lead us to suggest using the Cosine distance when employing Zest to select a suitable surrogate.
For completeness, we report all the correlation values for $\epsilon = 0.1$ and $N \in \{128, 64, 32\}$ in \Cref{tab:correlation_Ns}.

% UNIFIED CORRELATION TABLES
\begin{table*}
\centering
\footnotesize
\caption{Pearson correlation between transfer rate and Zest distance, for different values of $N \in \{128, 64, 32\}$, $\epsilon = 0.1$.}
\vskip 0.15in
\label{tab:correlation_Ns}
\begin{tabular}{r| cc | cc | cc}
\textbf{Model} & \multicolumn{2}{c}{$N=128$} & \multicolumn{2}{c}{$N=64$} & \multicolumn{2}{c}{$N=32$} \\
 &  \textbf{Corr - Cosine} & \textbf{Corr - $L_\infty$} &  \textbf{Corr - Cosine} & \textbf{Corr - $L_\infty$} &  \textbf{Corr - Cosine} & \textbf{Corr - $L_\infty$} \\
\midrule
\textbf{densenet121  }&  -0.830 & -0.435  &  -0.808 & -0.353  &  -0.793 & -0.392  \\
\textbf{densenet161  }&  -0.871 & -0.746  &  -0.850 & -0.593  &  -0.868 & -0.392  \\
\textbf{densenet169  }&  -0.814 & -0.641  &  -0.795 & -0.407  &  -0.767 & -0.534  \\
\textbf{googlenet    }&  -0.440 & -0.499  &  -0.437 & -0.469  &  -0.446 & -0.141  \\
\textbf{inception\_v3} &  -0.632 & -0.218  &  -0.590 & -0.245  &  -0.533 & -0.200 \\
\textbf{mobilenet\_v2} &  -0.111 & -0.527  &  -0.148 & -0.566  &  -0.176 & -0.111 \\
\textbf{resnet18     }&  -0.765 & -0.703  &  -0.732 & -0.726  &  -0.789 & -0.669  \\
\textbf{resnet34     }&  -0.734 & -0.513  &  -0.722 & -0.573  &  -0.709 & -0.788  \\
\textbf{resnet50     }&  -0.743 & -0.483  &  -0.698 & -0.443  &  -0.677 &  0.053  \\
\textbf{vgg11\_bn    } &   0.026 & -0.244  &   0.008 & -0.242  &  -0.031 & -0.269 \\
\textbf{vgg13\_bn    } &  -0.355 &  0.124  &  -0.371 & -0.424  &  -0.368 & -0.182 \\
\textbf{vgg16\_bn    } &  -0.724 & -0.157  &  -0.718 & -0.511  &  -0.757 & -0.721 \\
\textbf{vgg19\_bn    } &  -0.725 & -0.671  &  -0.700 & -0.534  &  -0.764 & -0.597 \\

\end{tabular}
\vskip -0.1in
\end{table*}

\subsection{Attacking a deployed model}

To empirically show the advantage in using the model-distance strategy when attacking a deployed classification system, we target an instance of a CIFAR-10 classifier generated through a machine learning as-a-service system (MLaaS).
This model is trained using exclusively automated tools, and neither its architecture nor its training parameters are known to us.
It achieves an area under the precision/recall curve of 0.93, and 88\% precision and 81.5\% recall for a confidence threshold value of 0.5.
Moreover, we can interact with it only through the APIs provided by the MLaaS platform.

We used the evasive samples computed with $\epsilon=0.1$, and show the results of this experiment, for both $N=128$ and $N=32$ (corresponding to 128,000 and 32,000 queries), in \Cref{mlaas_128v32_cosine} and \Cref{mlaas_128v32_linf}.
The closest 
Due to the known issue of numerical precision loss, when converting the adversarial examples generated locally to JPG files, we observe a reduction in the overall evasiveness of the samples. 
We also account for the fact that the 100 images chosen to compute the samples on, were not all correctly classified by the MLaaS model, which achieved 92\% accuracy on that subset, and therefore we only measure the accuracy difference after the attack.
Despite these limitations, we still observe a clear negative correlation, for both values of $N$, especially when using the Cosine metric.

\section{Discussion and Conclusion}

In this paper we proposed a simple and effective method for selecting suitable surrogate models to use when crafting adversarial examples to attack a black-box model via adversarial transfer. 
We showed that there is a strong negative correlation between the successful adversarial transfer rate between two models and their Zest distance, especially when evaluated with the Cosine metric.
We argue that this information can be leveraged to reduce the cost of an evasion campaign, as the price to compute the LIME signature of a victim model is paid only once, and can be re-used indefinitely.
Moreover, once a suitable proxy model has been identified, the adversary can craft arbitrarily many evasive samples without having to spend additional resources querying the victim model.

While the proposed method is a way to reduce adversarial cost, it is not designed to increase adversarial success per-se.
The adversarial examples are crafted with a powerful gradient-based method, but their effectiveness is bound by both the parameters chosen to craft the samples, and the transfer rate of the best proxy model the adversary has at their disposal.
This means that, if the adversary is incapable of acquiring an ample and varied portfolio of proxy models, and chose appropriate parameters for their adversarial examples crafting technique, using the proposed approach would not significantly help them.

While ML explainability methods are important for interpreting the predictions made by ML models, our work shows that these techniques might be weaponized by adversaries for launching attacks against ML. Understanding in more details the tradeoffs between ML explainability and robustness is an interesting topic for future work. 

\subsection{Ethical Considerations}
We are aware that the method we are proposing in this paper is an offensive technique, designed to reduce the attack cost of a potential malicious party.
This work has the purpose of exposing this possibility to the security community so it can be taken into account and used to better estimate adversarial costs when designing defenses in ML systems.
We also remind the reader that there are situations where ML evasion techniques can be directly beneficial to innocent people.
This is the case, for instance, of political activists and people living under dictatorial regimes, who may employ adversarial machine learning in an effort to evade oppressive systems such as facial recognition tools. 

% \section{Conclusions}

% \bibliographystyle{icml2022}
\bibliographystyle{IEEEtran}
\bibliography{prj_lemon}

\appendix
% \onecolumn
\section{Distances for Different Values of \texorpdfstring{$N$}{N}}\label{app:additional_dist}

% TABLE FOR N=128

\begin{table*}
\scriptsize
\centering
\caption{Zest L-inf and Cosine distance between all pairs of CIFAR-10 models, with $N = 128$.}
\label{tab:all_both_distances128}
\begin{tabular}{lr|rrrrrrrrrrrrr}
       &          & \VertTab{densenet121} & \VertTab{densenet161} & \VertTab{densenet169} & \VertTab{googlenet} & \VertTab{inception\_v3} & \VertTab{mobilenet\_v2} & \VertTab{resnet18} & \VertTab{resnet34} & \VertTab{resnet50} & \VertTab{vgg11\_bn} & \VertTab{vgg13\_bn} & \VertTab{vgg16\_bn} & \VertTab{vgg19\_bn} \\
\midrule
\multirow{13}{*}{\rotatebox{90}{$\mathbf{L_\infty}$}} & densenet121 &                   --- &                 2.526 &                 4.596 &               3.663 &                   3.567 &                   2.865 &              3.143 &              2.565 &              2.910 &               4.058 &               3.552 &               3.837 &               5.815 \\
       & densenet161 &                 2.526 &                   --- &                 2.478 &               4.076 &                   3.728 &                   3.231 &              3.022 &              3.680 &              3.064 &               3.674 &               3.513 &               4.184 &               5.868 \\
       & densenet169 &                 4.596 &                 2.478 &                   --- &               4.291 &                   3.475 &                   4.146 &              2.669 &              2.961 &              2.404 &               3.806 &               3.818 &               4.733 &               5.567 \\
       & googlenet &                 3.663 &                 4.076 &                 4.291 &                 --- &                   3.658 &                   3.607 &              5.145 &              4.868 &              3.763 &               4.266 &               3.565 &               3.946 &               5.478 \\
       & inception\_v3 &                 3.567 &                 3.728 &                 3.475 &               3.658 &                     --- &                   3.299 &              3.438 &              3.506 &              2.759 &               2.916 &               3.484 &               3.705 &               5.038 \\
       & mobilenet\_v2 &                 2.865 &                 3.231 &                 4.146 &               3.607 &                   3.299 &                     --- &              3.681 &              3.497 &              2.602 &               3.312 &               3.504 &               4.030 &               5.286 \\
       & resnet18 &                 3.143 &                 3.022 &                 2.669 &               5.145 &                   3.438 &                   3.681 &                --- &              2.890 &              2.866 &               3.453 &               3.476 &               3.766 &               5.601 \\
       & resnet34 &                 2.565 &                 3.680 &                 2.961 &               4.868 &                   3.506 &                   3.497 &              2.890 &                --- &              3.356 &               3.519 &               3.403 &               3.243 &               5.312 \\
       & resnet50 &                 2.910 &                 3.064 &                 2.404 &               3.763 &                   2.759 &                   2.602 &              2.866 &              3.356 &                --- &               3.320 &               3.383 &               3.613 &               4.863 \\
       & vgg11\_bn &                 4.058 &                 3.674 &                 3.806 &               4.266 &                   2.916 &                   3.312 &              3.453 &              3.519 &              3.320 &                 --- &               4.781 &               4.631 &               4.219 \\
       & vgg13\_bn &                 3.552 &                 3.513 &                 3.818 &               3.565 &                   3.484 &                   3.504 &              3.476 &              3.403 &              3.383 &               4.781 &                 --- &               2.687 &               4.085 \\
       & vgg16\_bn &                 3.837 &                 4.184 &                 4.733 &               3.946 &                   3.705 &                   4.030 &              3.766 &              3.243 &              3.613 &               4.631 &               2.687 &                 --- &               4.296 \\
       & vgg19\_bn &                 5.815 &                 5.868 &                 5.567 &               5.478 &                   5.038 &                   5.286 &              5.601 &              5.312 &              4.863 &               4.219 &               4.085 &               4.296 &                 --- \\
\midrule
\multirow{13}{*}{\rotatebox{90}{\textbf{Cosine}}} & densenet121 &                   --- &                 0.141 &                 0.141 &               0.262 &                   0.229 &                   0.201 &              0.180 &              0.177 &              0.145 &               0.285 &               0.267 &               0.283 &               0.335 \\
       & densenet161 &                 0.141 &                   --- &                 0.146 &               0.290 &                   0.232 &                   0.211 &              0.181 &              0.172 &              0.161 &               0.286 &               0.278 &               0.288 &               0.339 \\
       & densenet169 &                 0.141 &                 0.146 &                   --- &               0.289 &                   0.245 &                   0.204 &              0.191 &              0.183 &              0.144 &               0.309 &               0.290 &               0.297 &               0.343 \\
       & googlenet &                 0.262 &                 0.290 &                 0.289 &                 --- &                   0.236 &                   0.248 &              0.319 &              0.330 &              0.246 &               0.337 &               0.292 &               0.339 &               0.358 \\
       & inception\_v3 &                 0.229 &                 0.232 &                 0.245 &               0.236 &                     --- &                   0.192 &              0.256 &              0.265 &              0.219 &               0.319 &               0.271 &               0.304 &               0.328 \\
       & mobilenet\_v2 &                 0.201 &                 0.211 &                 0.204 &               0.248 &                   0.192 &                     --- &              0.214 &              0.218 &              0.182 &               0.288 &               0.258 &               0.263 &               0.303 \\
       & resnet18 &                 0.180 &                 0.181 &                 0.191 &               0.319 &                   0.256 &                   0.214 &                --- &              0.155 &              0.166 &               0.278 &               0.266 &               0.277 &               0.326 \\
       & resnet34 &                 0.177 &                 0.172 &                 0.183 &               0.330 &                   0.265 &                   0.218 &              0.155 &                --- &              0.170 &               0.281 &               0.281 &               0.274 &               0.330 \\
       & resnet50 &                 0.145 &                 0.161 &                 0.144 &               0.246 &                   0.219 &                   0.182 &              0.166 &              0.170 &                --- &               0.273 &               0.243 &               0.261 &               0.305 \\
       & vgg11\_bn &                 0.285 &                 0.286 &                 0.309 &               0.337 &                   0.319 &                   0.288 &              0.278 &              0.281 &              0.273 &                 --- &               0.160 &               0.180 &               0.238 \\
       & vgg13\_bn &                 0.267 &                 0.278 &                 0.290 &               0.292 &                   0.271 &                   0.258 &              0.266 &              0.281 &              0.243 &               0.160 &                 --- &               0.130 &               0.177 \\
       & vgg16\_bn &                 0.283 &                 0.288 &                 0.297 &               0.339 &                   0.304 &                   0.263 &              0.277 &              0.274 &              0.261 &               0.180 &               0.130 &                 --- &               0.162 \\
       & vgg19\_bn &                 0.335 &                 0.339 &                 0.343 &               0.358 &                   0.328 &                   0.303 &              0.326 &              0.330 &              0.305 &               0.238 &               0.177 &               0.162 &                 --- \\
\bottomrule
\end{tabular}
\end{table*}

% TABLE FOR N = 64

\begin{table*}[ht]
\centering
\scriptsize
\caption{Zest L-inf and Cosine distance between all pairs of CIFAR-10 models, with $N = 64$.}
\label{tab:all_both_distances64}
\begin{tabular}{lr|rrrrrrrrrrrrr}
       &          & \VertTab{densenet121} & \VertTab{densenet161} & \VertTab{densenet169} & \VertTab{googlenet} & \VertTab{inception\_v3} & \VertTab{mobilenet\_v2} & \VertTab{resnet18} & \VertTab{resnet34} & \VertTab{resnet50} & \VertTab{vgg11\_bn} & \VertTab{vgg13\_bn} & \VertTab{vgg16\_bn} & \VertTab{vgg19\_bn} \\
\midrule
\multirow{13}{*}{\rotatebox{90}{$\mathbf{L_\infty}$}} & densenet121 &                   --- &                 2.526 &                 4.596 &               3.663 &                   2.741 &                   2.865 &              3.143 &              2.327 &              2.910 &               4.058 &               3.552 &               3.837 &               7.165 \\
       & densenet161 &                 2.526 &                   --- &                 2.365 &               4.076 &                   2.710 &                   2.597 &              2.700 &              2.753 &              3.064 &               2.964 &               3.513 &               3.487 &               6.309 \\
       & densenet169 &                 4.596 &                 2.365 &                   --- &               3.775 &                   2.605 &                   4.146 &              2.646 &              2.961 &              2.404 &               2.989 &               4.181 &               3.420 &               5.496 \\
       & googlenet &                 3.663 &                 4.076 &                 3.775 &                 --- &                   3.658 &                   3.607 &              5.145 &              4.868 &              3.763 &               4.266 &               3.565 &               4.071 &               5.478 \\
       & inception\_v3 &                 2.741 &                 2.710 &                 2.605 &               3.658 &                     --- &                   3.101 &              3.173 &              3.506 &              2.623 &               2.916 &               3.484 &               3.705 &               4.854 \\
       & mobilenet\_v2 &                 2.865 &                 2.597 &                 4.146 &               3.607 &                   3.101 &                     --- &              3.681 &              2.699 &              2.538 &               3.312 &               3.504 &               4.144 &               5.716 \\
       & resnet18 &                 3.143 &                 2.700 &                 2.646 &               5.145 &                   3.173 &                   3.681 &                --- &              2.890 &              2.866 &               2.951 &               3.476 &               3.766 &               4.843 \\
       & resnet34 &                 2.327 &                 2.753 &                 2.961 &               4.868 &                   3.506 &                   2.699 &              2.890 &                --- &              3.356 &               2.949 &               3.403 &               3.243 &               5.297 \\
       & resnet50 &                 2.910 &                 3.064 &                 2.404 &               3.763 &                   2.623 &                   2.538 &              2.866 &              3.356 &                --- &               3.320 &               3.383 &               3.613 &               4.692 \\
       & vgg11\_bn &                 4.058 &                 2.964 &                 2.989 &               4.266 &                   2.916 &                   3.312 &              2.951 &              2.949 &              3.320 &                 --- &               3.063 &               2.570 &               4.219 \\
       & vgg13\_bn &                 3.552 &                 3.513 &                 4.181 &               3.565 &                   3.484 &                   3.504 &              3.476 &              3.403 &              3.383 &               3.063 &                 --- &               2.687 &               4.085 \\
       & vgg16\_bn &                 3.837 &                 3.487 &                 3.420 &               4.071 &                   3.705 &                   4.144 &              3.766 &              3.243 &              3.613 &               2.570 &               2.687 &                 --- &               2.921 \\
       & vgg19\_bn &                 7.165 &                 6.309 &                 5.496 &               5.478 &                   4.854 &                   5.716 &              4.843 &              5.297 &              4.692 &               4.219 &               4.085 &               2.921 &                 --- \\
\midrule
\multirow{13}{*}{\rotatebox{90}{\textbf{Cosine}}} & densenet121 &                   --- &                 0.143 &                 0.156 &               0.270 &                   0.221 &                   0.212 &              0.189 &              0.186 &              0.158 &               0.291 &               0.264 &               0.279 &               0.345 \\
       & densenet161 &                 0.143 &                   --- &                 0.153 &               0.301 &                   0.225 &                   0.213 &              0.187 &              0.178 &              0.170 &               0.282 &               0.274 &               0.285 &               0.350 \\
       & densenet169 &                 0.156 &                 0.153 &                   --- &               0.299 &                   0.243 &                   0.211 &              0.199 &              0.187 &              0.153 &               0.322 &               0.292 &               0.294 &               0.351 \\
       & googlenet &                 0.270 &                 0.301 &                 0.299 &                 --- &                   0.240 &                   0.255 &              0.337 &              0.350 &              0.255 &               0.344 &               0.298 &               0.342 &               0.369 \\
       & inception\_v3 &                 0.221 &                 0.225 &                 0.243 &               0.240 &                     --- &                   0.196 &              0.250 &              0.269 &              0.219 &               0.320 &               0.270 &               0.300 &               0.335 \\
       & mobilenet\_v2 &                 0.212 &                 0.213 &                 0.211 &               0.255 &                   0.196 &                     --- &              0.211 &              0.222 &              0.177 &               0.293 &               0.260 &               0.261 &               0.315 \\
       & resnet18 &                 0.189 &                 0.187 &                 0.199 &               0.337 &                   0.250 &                   0.211 &                --- &              0.166 &              0.167 &               0.277 &               0.259 &               0.273 &               0.328 \\
       & resnet34 &                 0.186 &                 0.178 &                 0.187 &               0.350 &                   0.269 &                   0.222 &              0.166 &                --- &              0.188 &               0.285 &               0.284 &               0.280 &               0.344 \\
       & resnet50 &                 0.158 &                 0.170 &                 0.153 &               0.255 &                   0.219 &                   0.177 &              0.167 &              0.188 &                --- &               0.278 &               0.247 &               0.260 &               0.312 \\
       & vgg11\_bn &                 0.291 &                 0.282 &                 0.322 &               0.344 &                   0.320 &                   0.293 &              0.277 &              0.285 &              0.278 &                 --- &               0.160 &               0.179 &               0.252 \\
       & vgg13\_bn &                 0.264 &                 0.274 &                 0.292 &               0.298 &                   0.270 &                   0.260 &              0.259 &              0.284 &              0.247 &               0.160 &                 --- &               0.131 &               0.186 \\
       & vgg16\_bn &                 0.279 &                 0.285 &                 0.294 &               0.342 &                   0.300 &                   0.261 &              0.273 &              0.280 &              0.260 &               0.179 &               0.131 &                 --- &               0.160 \\
       & vgg19\_bn &                 0.345 &                 0.350 &                 0.351 &               0.369 &                   0.335 &                   0.315 &              0.328 &              0.344 &              0.312 &               0.252 &               0.186 &               0.160 &                 --- \\
\bottomrule
\end{tabular}
\end{table*}

% TABLE FOR N = 32

\begin{table*}[ht]
\centering
\scriptsize
\caption{Zest L-inf and Cosine distance between all pairs of CIFAR-10 models, with $N = 32$.}
\label{tab:all_both_distances32}
\begin{tabular}{lr|rrrrrrrrrrrrr}
       &          & \VertTab{densenet121} & \VertTab{densenet161} & \VertTab{densenet169} & \VertTab{googlenet} & \VertTab{inception\_v3} & \VertTab{mobilenet\_v2} & \VertTab{resnet18} & \VertTab{resnet34} & \VertTab{resnet50} & \VertTab{vgg11\_bn} & \VertTab{vgg13\_bn} & \VertTab{vgg16\_bn} & \VertTab{vgg19\_bn} \\
\midrule
\multirow{13}{*}{\rotatebox{90}{$\mathbf{L_\infty}$}} & densenet121 &                   --- &                 2.526 &                 2.081 &               3.115 &                   2.741 &                   2.809 &              2.763 &              2.327 &              4.305 &               2.864 &               3.552 &               3.654 &               3.432 \\
       & densenet161 &                 2.526 &                   --- &                 2.365 &               2.748 &                   2.710 &                   2.597 &              2.268 &              2.753 &              3.064 &               4.613 &               4.076 &               3.487 &               3.469 \\
       & densenet169 &                 2.081 &                 2.365 &                   --- &               2.997 &                   2.605 &                   2.067 &              2.570 &              2.961 &              2.131 &               2.935 &               3.465 &               2.976 &               3.720 \\
       & googlenet &                 3.115 &                 2.748 &                 2.997 &                 --- &                   2.684 &                   2.948 &              2.754 &              3.270 &              2.522 &               3.339 &               4.020 &               3.609 &               3.776 \\
       & inception\_v3 &                 2.741 &                 2.710 &                 2.605 &               2.684 &                     --- &                   3.101 &              2.989 &              3.942 &              2.500 &               3.588 &               3.484 &               4.168 &               3.540 \\
       & mobilenet\_v2 &                 2.809 &                 2.597 &                 2.067 &               2.948 &                   3.101 &                     --- &              2.594 &              3.178 &              2.322 &               2.939 &               4.067 &               3.344 &               3.758 \\
       & resnet18 &                 2.763 &                 2.268 &                 2.570 &               2.754 &                   2.989 &                   2.594 &                --- &              2.062 &              2.866 &               2.794 &               3.476 &               2.986 &               3.730 \\
       & resnet34 &                 2.327 &                 2.753 &                 2.961 &               3.270 &                   3.942 &                   3.178 &              2.062 &                --- &              3.356 &               2.949 &               3.403 &               3.190 &               3.827 \\
       & resnet50 &                 4.305 &                 3.064 &                 2.131 &               2.522 &                   2.500 &                   2.322 &              2.866 &              3.356 &                --- &               2.737 &               3.383 &               3.613 &               3.486 \\
       & vgg11\_bn &                 2.864 &                 4.613 &                 2.935 &               3.339 &                   3.588 &                   2.939 &              2.794 &              2.949 &              2.737 &                 --- &               1.934 &               2.570 &               2.590 \\
       & vgg13\_bn &                 3.552 &                 4.076 &                 3.465 &               4.020 &                   3.484 &                   4.067 &              3.476 &              3.403 &              3.383 &               1.934 &                 --- &               2.241 &               2.161 \\
       & vgg16\_bn &                 3.654 &                 3.487 &                 2.976 &               3.609 &                   4.168 &                   3.344 &              2.986 &              3.190 &              3.613 &               2.570 &               2.241 &                 --- &               2.280 \\
       & vgg19\_bn &                 3.432 &                 3.469 &                 3.720 &               3.776 &                   3.540 &                   3.758 &              3.730 &              3.827 &              3.486 &               2.590 &               2.161 &               2.280 &                 --- \\
\midrule
\multirow{13}{*}{\rotatebox{90}{\textbf{Cosine}}} & densenet121 &                   --- &                 0.160 &                 0.132 &               0.266 &                   0.223 &                   0.201 &              0.175 &              0.189 &              0.141 &               0.285 &               0.251 &               0.285 &               0.336 \\
       & densenet161 &                 0.160 &                   --- &                 0.165 &               0.323 &                   0.250 &                   0.221 &              0.171 &              0.190 &              0.187 &               0.289 &               0.281 &               0.292 &               0.348 \\
       & densenet169 &                 0.132 &                 0.165 &                   --- &               0.307 &                   0.238 &                   0.188 &              0.179 &              0.178 &              0.153 &               0.306 &               0.285 &               0.294 &               0.345 \\
       & googlenet &                 0.266 &                 0.323 &                 0.307 &                 --- &                   0.251 &                   0.255 &              0.339 &              0.359 &              0.253 &               0.341 &               0.296 &               0.334 &               0.357 \\
       & inception\_v3 &                 0.223 &                 0.250 &                 0.238 &               0.251 &                     --- &                   0.213 &              0.256 &              0.277 &              0.222 &               0.321 &               0.272 &               0.305 &               0.344 \\
       & mobilenet\_v2 &                 0.201 &                 0.221 &                 0.188 &               0.255 &                   0.213 &                     --- &              0.201 &              0.201 &              0.176 &               0.280 &               0.260 &               0.248 &               0.298 \\
       & resnet18 &                 0.175 &                 0.171 &                 0.179 &               0.339 &                   0.256 &                   0.201 &                --- &              0.146 &              0.179 &               0.257 &               0.253 &               0.268 &               0.319 \\
       & resnet34 &                 0.189 &                 0.190 &                 0.178 &               0.359 &                   0.277 &                   0.201 &              0.146 &                --- &              0.196 &               0.288 &               0.279 &               0.279 &               0.326 \\
       & resnet50 &                 0.141 &                 0.187 &                 0.153 &               0.253 &                   0.222 &                   0.176 &              0.179 &              0.196 &                --- &               0.270 &               0.240 &               0.270 &               0.310 \\
       & vgg11\_bn &                 0.285 &                 0.289 &                 0.306 &               0.341 &                   0.321 &                   0.280 &              0.257 &              0.288 &              0.270 &                 --- &               0.152 &               0.170 &               0.235 \\
       & vgg13\_bn &                 0.251 &                 0.281 &                 0.285 &               0.296 &                   0.272 &                   0.260 &              0.253 &              0.279 &              0.240 &               0.152 &                 --- &               0.125 &               0.177 \\
       & vgg16\_bn &                 0.285 &                 0.292 &                 0.294 &               0.334 &                   0.305 &                   0.248 &              0.268 &              0.279 &              0.270 &               0.170 &               0.125 &                 --- &               0.160 \\
       & vgg19\_bn &                 0.336 &                 0.348 &                 0.345 &               0.357 &                   0.344 &                   0.298 &              0.319 &              0.326 &              0.310 &               0.235 &               0.177 &               0.160 &                 --- \\
\bottomrule
\end{tabular}
\end{table*}

Here, we report the Zest distances computed for all possible pairs of the 13 CIFAR-10 models we analyzed.
The distances are shown for both the $L_\infty$ and Cosine metrics, as those metrics where the ones which appeared to better correlate with adversarial transfer, with the best being Cosine.
We also report these results for different value of the number of representative samples used to compute the LIME models, $N$, with \Cref{tab:all_both_distances128}, \Cref{tab:all_both_distances64} and \Cref{tab:all_both_distances32} showing values for $N=128, 64, 32$ respectively.
Lower values of $N$ are interesting from an adversarial perspective, as they reduce the number of queries to the victim model necessary to compute the LIME representation, reducing cost further.

\end{document}